\documentclass[useAMS,usenatbib]{mn2e}
\usepackage{mycommands}
\usepackage{verbatim}
\usepackage{graphicx}
\usepackage{subfigure}
\usepackage[percent]{overpic}
\usepackage{esint}
\usepackage{url}
\title[Pulsar Dissipation, Energy Transer, and Spindown Luminosity]{Dissipation, Energy Transfer, and Spindown Luminosity in 2.5D PIC Simulations of the Pulsar Magnetosphere}
\author[Mikhail A. Belyaev]{Mikhail A. Belyaev$^{1}$\thanks{email: mbelyaev@berkeley.edu} \\
$^{1}$Astronomy Department, University of California, Berkeley, CA 94720}

\begin{document}

\maketitle

\begin{abstract}
We perform 2.5D axisymmetric simulations of the pulsar magnetosphere (aligned dipole rotator) using the charge conservative, relativistic, electromagnetic particle in cell code PICsar. Particle in cell codes are a powerful tool to use for studying the pulsar magnetosphere, because they can handle the force-free and vacuum limits and provide a self-consistent treatment of magnetic reconnection.  In the limit of dense plasma throughout the magnetosphere, our solutions are everywhere in the force-free regime except for dissipative regions at the polar caps, in the current layers, and at the Y-point. These dissipative regions arise self-consistently, since we do not have any explicit dissipation in the code. A minimum of $ \approx 15-20\%$ of the electromagnetic spindown luminosity is transferred to the particles inside 5 light cylinder radii. However, the particles can carry as much as $\gtrsim 50 \%$ of the spindown luminosity if there is insufficient plasma in the outer magnetosphere to screen the component of electric field parallel to the magnetic field. In reality, the component of the spindown luminosity carried by the particles could be radiated as gamma rays, but high-frequency synchrotron emission would need to be implemented as a sub-grid process in our simulations and is not present for the current suite of runs. The value of the spindown luminosity in our simulations is within $\approx 10\%$ of the force-free value, and the structure of the electromagnetic fields in the magnetosphere is on the whole consistent with the force-free model.
\end{abstract}
\date{}

\section{Introduction}
\label{intro}

Observations have yielded a wealth of gamma ray pulsars and have demonstrated pulsars to be the dominant source of GeV gamma rays in the Milky Way \citep{Fermi}. Indeed, pulsed periodic emission from neutron stars has been observed across a range of wavelength bands from radio to gamma rays. Nevertheless, a comprehensive physical model of pulsars and pulsar emission remains elusive.

Over the past two decades, computational studies of the pulsar magnetosphere, starting with direct integration of the time-independent axisymmetric ``pulsar equation" \citep{Contopoulos,OguraKojima,Gruzinov,Timokhin_forcefree} and more recently time-dependent ideal and resistive force-free and MHD simulations \citep{Komissarov,Spitkovsky,McKinney,CK,LiSpitkovsky,Kalapotharakos,PetriFF,Parfreyetal,Tchekhovskoy} have yielded the basic underlying morphology of the pulsar magnetosphere as well as a formula for the spindown luminosity at arbitrary inclination angles. 

Knowledge of pulsar magnetosphere structure can be used to discriminate between potential sites of particle acceleration and high-energy emission and to fit observed pulsar light curves \citep{DyksRudak,BaiSpitkovsky,RomaniWatters,Kalapotharakos_lightcurve}.

A drawback of force-free and MHD simulations are that they assume the magnetosphere is filled with a plasma that is dense enough to short out the accelerating component of the electric field on the particles (modulo a resistivity in the case of resistive simulations). However, locations where the force-free approximation breaks down and the plasma is not dense enough to screen the accelerating electric field (vacuum gaps) or where the magnetic field vanishes (equatorial current sheet and Y-point) are the sites of particle acceleration and emission. Thus, it is necessary to go beyond force-free and MHD models in order to self-consistently model particle acceleration and emission in pulsar magnetospheres. 

Recently, there has been a flurry of activity in the area of global simulations of the pulsar magnetosphere using the particle in cell (PIC) method in axisymmetry \citep{chenbeloborodov,CeruttiSpitkovsky} and in full 3D \citep{PhilippovSpitkovsky,PhilippovSpitkovsky1}. One important advantage PIC methods have over force-free and MHD techniques is that they are able to self-consistently simulate vacuum gaps and current sheets, where the force-free conditions $\bfE \cdot \bfB = 0$ and $\bfE \cdot \bfJ = 0$ are violated. Thus, it is possible to carry out first principles simulations of the particle acceleration and emission in the magnetosphere using PIC simulations. 

For example, \citet{TimokhinArons} used 1D PIC simulations local to the pulsar polar cap to study low altitude particle acceleration and pair production, and \citet{ZenitaniHoshino2007,ZenitaniHoshino2008} used PIC simulations to model the microphysics of magnetic reconnection, particle acceleration, and plasma instabilities in relativistic current sheets.

This paper focuses on simulations of the axisymmetric pulsar magnetosphere using the PIC code {\it PICsar} \citep{BelyaevPIC}. We begin by discussing the code and the setup of our simulations in \S\ref{methods}, and we present our results in \S \ref{results}. 

When a dense plasma is present throughout the magnetosphere, our results are consistent with the canonical force-free model for the structure of the fields in the magnetosphere; they also agree with the value of the force-free spindown luminosity to within $10 \%$.  However, we find that even in the limit of high plasma density, there exist regions of particle acceleration in the magnetosphere, where $\bfE \cdot \bfJ \ne 0$. As a result, at least $15-20 \%$ of the spindown luminosity within 5 light cylinder radii is carried by particles, and this fraction can be as high as $\gtrsim 50\%$ if there is insufficient plasma in the outer magnetosphere to short out the component of electric field parallel to the magnetic field. Because high-frequency synchrotron radiation is a sub-grid process that is not captured by our current suite of simulations, the component of spindown luminosity carried by the particles could potentially be reradiated as high-energy emission. 

\section{Numerical Methods and Simulation Setup}
\label{methods}

\subsection{PIC Methods}

For our simulations we use the 2.5D axisymmetric, electromagnetic, relativistic, charge conservative PIC code {\it PICsar}. It is based on the 3D electromagnetic, relativistic, charge conservative PIC code {\it TRISTAN} \citep{TRISTAN}, but has been parallelized using MPI and modified specifically for pulsar simulations. {\it PICsar} has been extensively tested, and the numerical methods are thoroughly documented in \citet{BelyaevPIC}, so we provide only a brief overview of them here. 

{\it PICsar} solves the two time-dependent Maxwell's equations in integral form on a curvilinear grid using standard FDTD techniques \citep{Yee,THREDS,fdtdbook}. The two time-independent Maxwell's equations (Poisson's equation and zero divergence for the magnetic field) are satisfied to machine precision for all time as long as they are satisfied initially. This is by virtue of the FDTD method coupled with a charge conservative deposition scheme for the current \citep{VillasenorBuneman}.

For the particle mover, we prefer the Vay algorithm \citep{Vay} over the classical Boris algorithm \citep{Boris}, since the Vay algorithm reproduces the correct particle $E \times B$ drift even when the gyrofrequency and Larmor radius of a particle are unresolved \citep{Vay, BelyaevPIC}. Since the $E \times B$ drift is the dominant particle drift in the magnetosphere, this is an important feature in a global pulsar PIC simulation, for which the magnetic field varies by orders of magnitude across the simulation domain.

Just as in the original TRISTAN code, we use digital charge filtering to reduce the high-frequency shot noise on the grid due to finite number particle statistics. The filter weights are the same as those of the ``1-2-1" digital filter used in TRISTAN, except near the polar axis, where the weights must be modified to ensure rigorous charge conservation \citep{BelyaevPIC}. 

\subsection{Coordinate Grid}
\label{coordgrid}
The particular choice of grid we use for our simulations is logarithmic in the radial direction and ``equal area" in the meridional direction. The coordinates are given in terms of the usual spherical $r - \theta$ coordinates as
\begin{align}
\xi &= r_* \ln(r/r_*) \\
\theta_A &= -\cos(\theta) \nn
\end{align}
where $\xi$ is the radial coordinate, $\theta_A$ is the meridional coordinate, and $r_*$ is the inner radial boundary of the simulation domain, which coincides with the surface of the neutron star. An image of the grid lines for the $\xi$-$\theta_A$ coordinate system and a discussion of its advantages over the $(r, \theta)$ and $(\xi, \theta)$ coordinate systems for PIC simulations are given in \citet{BelyaevPIC}.

\subsection{Boundary Conditions}
\label{bcsec}
The inner radial boundary of the simulation domain is the surface of the neutron star and is a perfectly conducting boundary. The outer radial boundary of the simulation domain is a damping layer, which damps incident EM waves \citep{fdtdbook,BelyaevPIC}. However, we put the outer boundary sufficiently far out that any signal propagating at the speed of light from the surface of the neutron star does not have time to propagate to the outer boundary and back during the simulation. On the polar axis, we use a reflecting boundary condition for the particles and a special update for the EM fields derived using the integral form of the time-dependent Maxwell equations \citep{THREDS,BelyaevPIC}.

\subsection{Initial Conditions and Field Decomposition}
The initial condition for our simulations is a neutron star rotating in a background magnetic field for which the magnetic and spin axes are aligned. The initial magnetic field is assumed to be a dipole field, and an electric field is induced by the rotation of the neutron star (assumed to be a perfect spherical conductor). The initial vacuum electric and magnetic fields outside the neutron star in spherical coordinates are given by \citep{MichelLi}:
\begin{align}
B_r &= 2B_* \left(\frac{r_*}{r}\right)^3\cos \theta,  &E_r &=  B_* \left(\frac{r_*^5}{r^4R_\text{lc}}\right)(1- 3\cos^2 \theta)  \\
B_\theta &= B_* \left(\frac{r_*}{r}\right)^3\sin \theta,  &E_\theta &=-B_*\left(\frac{r_*^5}{r^4R_\text{lc}}\right) \sin 2\theta \nn \\
B_\phi &= 0, &E_\phi &= 0, \nn
\end{align}
where $r_*$ is the neutron star radius, $B_*$ is the surface magnetic field at the equator, and $R_\text{lc} \equiv c/\Omega_*$ is the light cylinder radius.

The vacuum fields are not represented on the numerical grid directly. Instead, a decomposition is applied to the electric and magnetic fields so that the total field is the sum of the analytic vacuum fields and the fields on the numerical grid \citep{Spitkovskythesis,BelyaevPIC}:
\begin{align}
B_\text{tot} &= B_\text{vac} + B_\text{grid} \\
E_\text{tot} &= E_\text{vac} + E_\text{grid} \nn
\end{align}
Thus, the electric and magnetic fields on the grid are initially zero and are created by the motion of particles in the simulation, which deposit currents to the grid (i.e. act as a source term in Maxwell's equations). The total field is used when accelerating particles during the mover step and when calculating how many particles to inject into the simulation domain.

To avoid a large initial transient, we linearly spin up the star on a timescale of half a rotation period \citep{Spitkovskythesis,BelyaevPIC}. After the initial spinup phase, the rotation rate of the star is fixed for the duration of the simulation.

\subsection{Particle Injection}
Initially there are {\it no} particles in the simulation domain in any of our simulations. Particles are injected into and deleted from the simulation in a charge-conservative fashion as it proceeds. All of the particles in our simulation have the same magnitude of charge and the same mass, so we are modeling an electron-positron plasma. 

Particles are created in oppositely charged pairs that are injected into the simulation at the same location, meaning we never have to solve Poisson's equation, due to the charge conservative nature of the code. 

The velocity of the injected particles is set to be the local $E \times B$ drift velocity, $\bfv_{inj} = c \bfE \times \bfB/B^2$, and we only inject a particle if $|v_{inj}/c| < 1$.  The latter condition is satisfied throughout the simulation domain except in the center of the equatorial current sheet. 

\subsubsection{Surface Emission of Particles}
\label{surfemitsec}
Because the neutron star in our simulations is modeled as a perfectly conducting sphere rotating in a dipole magnetic field, it must have a surface charge in vacuum to ensure the EM force vanishes inside the neutron star \citep{MichelLi}. We assume the EM fields at the surface of the neutron star are so strong that it has zero work function, so it cannot hang on to its surface charge.

The surface charge on a magnetized, rotating, conducting sphere (in Gaussian units) is given by the expression 
\begin{align}
\label{surfcharge}
4 \pi \sigma = \frac{\bfE \cdot \bfB}{B_r},
\end{align}
where $\bfE \cdot \bfB/B_r$ is evaluated on the surface of the conductor. To model surface emission, we locally inject the surface charge into the simulation domain in the form of particles.

Our exact ansatz for the local surface charge injection rate is given by the formula
\begin{align}
\label{surfchargeansatz}
4 \pi \dot{N} = \frac{f_\text{surf}}{q}\frac{|\bfE \cdot \bfB|}{B} dA,
\end{align}
where $\dot{N}$ is the number of pairs released (at rest) per timestep into the first cell above the surface of the conductor, $q$ is the charge of a positive particle (charge of a negative particle is $-q$ by assumption), $dA$ is the area element of the cell on the surface of the star, and $f_\text{surf} \lesssim 1$ determines the rate of surface charge injection. Since the charge is released in pairs at rest, we point out that the injected particle with the same sign of charge as the surface charge is accelerated into the simulation domain, whereas the injected particle with the opposite sign of charge falls into the star and is deleted.

The ansatz for releasing surface charge into the simulation domain (equation [\ref{surfchargeansatz}]) is modified in two significant ways relative the analytical formula (equation [\ref{surfcharge}]) to ensure numerical stability. The first modification is to only release a fraction of the surface charge, $f_\text{surf}$, in pairs at each timestep. The second modification is to replace the radial magnetic field, $B_r$, with the total magnetic field, $B$, in the denominator. This is because $B_r = 0$ at the equator for a dipole magnetic field, so small oscillations in the value of $\bfE \cdot \bfB$ due to noise can lead to numerical instability close to the equator. The effect of both of these modifications is negligible as long as the injection is much faster than the rate at which the surface charge is regenerated.

\subsubsection{Volume Injection of Particles}
\label{volinjsec}
Processes have been proposed that are capable of injecting plasma into magnetospheric gaps at large distances such as high-altitude pair production in a slot gap or outer gap \citep{ChengHoRuderman,AronsScharlemann,HardingMuslimov} and reconnection at the Y-point \citep{Arons_review}. Rather than modeling these microphysical processes directly, which is challenging in a global simulation due to the short timescales involved, we use a simplified but general ansatz for particle injection into the outer magnetosphere. In Gaussian units, this ansatz is
\begin{align}
\label{chargeinj}
4 \pi \dot{N} = \frac{f_\text{vol}(r)}{q r} \frac{|\bfE \cdot \bfB|}{B} dV,
\end{align}
where $\dot{N}$ is the number of pairs injected, $r$ is the spherical radius, $q$ is the charge of a positive particle, $dV$ is the volume element into which we are injecting, and $f_\text{vol}$ is a function of the radius that determines the rate of injection. 

The ansatz (\ref{chargeinj}) will drive the value of $\bfE \cdot \bfB$ to zero in regions where $f_\text{vol}(r) \sim 1$.
The form of the ansatz for volume charge injection is motivated by the ansatz for surface charge injection (equation [\ref{surfchargeansatz}]), and the extra factor of $1/r$ is necessary to make the units work. 

As a crude approximation to pair production, we assume that the function $f_\text{vol}(r)$ is constant with value $f_\text{vol}$ inside of a characteristic radius $r = r_\text{inj}$ and is zero outside of it. Thus, we use a tophat function for $f_\text{vol}(r)$:
\begin{align}
\label{fvoldef}
f_\text{vol}(r) = \left\{
     \begin{array}{lr}
       f_\text{vol} & : r \le r_{inj}\\
       0 & : r > r_{inj} 
     \end{array}
   \right.
\end{align}

\subsection{Simulation Setup and Parameters}
\begin{table}
\caption{Simulation parameters.} 
\label{simtable}
\centering
\begin{tabular}{|c|c|c|c|c|} 
\hline
label & $f_\text{vol}$ & $r_\text{inj}/R_\text{lc}$ & $t_\text{inj}/P_*$ \\
\hline
fvol\_.25\_rinj\_1.25 & .25 & $1.25$ & $10$  \\
fvol\_.5\_rinj\_1.25 & .5 & $1.25$ & $10$  \\
fvol\_1\_rinj\_1.25 & 1 & $1.25$ & $10$  \\
fvol\_.5\_rinj\_.9 & .5 & $.9$ & $10$ \\
injvol\_turnoff & .5 & $1.25$ & $3$ \\
\hline
\end{tabular}
\end{table}

In order for pulsar PIC simulations to be astrophysically relevant, they must have a high magnetization parameter. The magnetization is defined as
\begin{align}
\label{magnetization}
\sigma \equiv \frac{B^2}{4 \pi (\gamma_+ n_+ + \gamma_- n_-) m c^2},
\end{align}
where $\gamma_\pm n_\pm m c^2$ is the energy density of positrons/electrons. When $\sigma \gg 1$, the electromagnetic energy dominates the energy in the particles, particle inertia is negligible, and the force-free regime applies. Our simulations have a characteristic magnetization of $\sigma \sim 10^2-10^4$ at the light cylinder. However, the magnetization drops by orders of magnitude in the vicinity of the Y-point and current sheets such that $\sigma \lesssim 1$ there. 

Although we do not present the results here, we have checked that the exact value of the magnetic field we choose for our pulsar simulations does not affect the solution, as long as the characteristic magnetization at the light cylinder is high.

A second important physical parameter is the ratio of the light cylinder radius to the neutron star radius $R_\text{lc}/r_* = c/\Omega_* r_*$. Millisecond pulsars have $R_\text{lc}/r_* \sim 10$. However, in order to be able to run for $10 P_*$ ($P_* \equiv 2\pi/\Omega_*$ equals one orbital period), we use a smaller value of the light cylinder radius, such that  $R_\text{lc}/r_* = 4$ in all of our simulations. 

As mentioned in \S \ref{bcsec}, the outer edge of our simulation domain is placed far enough in radius that a signal propagating at the speed of light from the surface of the neutron star cannot make it to the outer boundary and back during the course of the simulation ($10 P_*$). The outer radius of our simulation domain is at $r_\text{out} \approx 148 r_* \approx 37 R_\text{lc}$ in all of our simulations. Placing the outer boundary this far out allows us to be completely certain that the outer boundary conditions do not influence our results. 

Other numerical parameters common to all of our simulations are as follows. The dimensions of the grid are $N_r \times N_\theta = 1280 \times 512$ cells (the cells have uniform spacing in the $(\xi, \theta_A)$ coordinate system), and the star has an effective radius of 256 cells (radial cell size is taken to be that of a cell at the surface of a star at the equator). The number of particles in all of our simulations in steady state is $N_\text{tot} \approx 3 \times 10^8$. 

For surface emission (\S \ref{surfemitsec}), we set the surface injection rate to $f_\text{surf} = .05$ in all of our simulations. This means that the characteristic time to release surface charge from the neutron star is $\approx 20$ timesteps, which is significantly shorter than the characteristic light-crossing time $r_*/c \approx 430$ timesteps. We have checked that the simulation results are essentially independent of the value of $f_\text{surf}$ as long as $1/f_\text{surf} \ll r_*/c$. However, setting $f_\text{surf} \gtrsim .1$ results in virtual cathode oscillations at the polar caps that become increasingly violent as $f_\text{surf}$ is increased further. 

The parameters which vary from simulation to simulation are given in Table \ref{simtable}. In our simulations, we vary the two parameters that characterize volume injection, $r_\text{inj}$ and $f_\text{vol}$ (equation [\ref{fvoldef}])). We also consider the effect of initializing a force-free steady state by turning on volume injection for a time $t_\text{inj}$ and then turning it off. Since all of our simulations have a duration of $10P_*$, those simulations, which have $t_\text{inj}/P_* = 10$ have volume injection turned on for the duration of the simulation. In simulation injvol\_turnoff, which has $t_\text{inj}/P_* = 3$, we start with volume injection turned on and turn it off after three rotation periods of the neutron star. Surface injection is turned on for the duration of the simulation in all simulations.

\section{Simulation Results}
\label{results}

\begin{figure*}
\centering
\subfigure{\begin{overpic}
		  [width=.3275\textwidth]{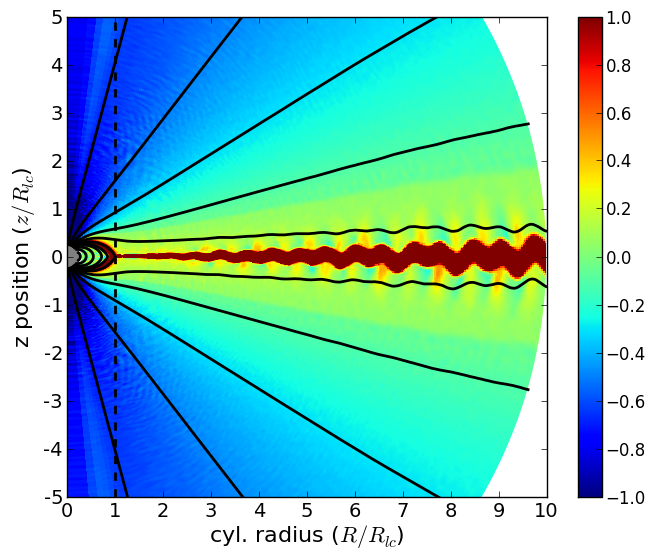}
		  \put(77,77){\large \bf a)}
		  \put(12,85){\large \bf meridional current}
		  \end{overpic}}
\subfigure{\begin{overpic}
		  [width=.32\textwidth]{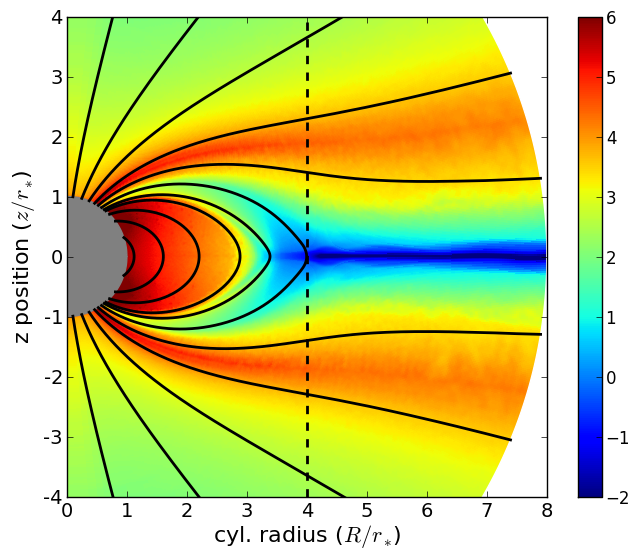}
		  \put(79,79){\large \bf b)}
		  \put(12,87){\large \bf log magnetization}
		  \end{overpic}}
\subfigure{\begin{overpic}
		  [width=.32\textwidth]{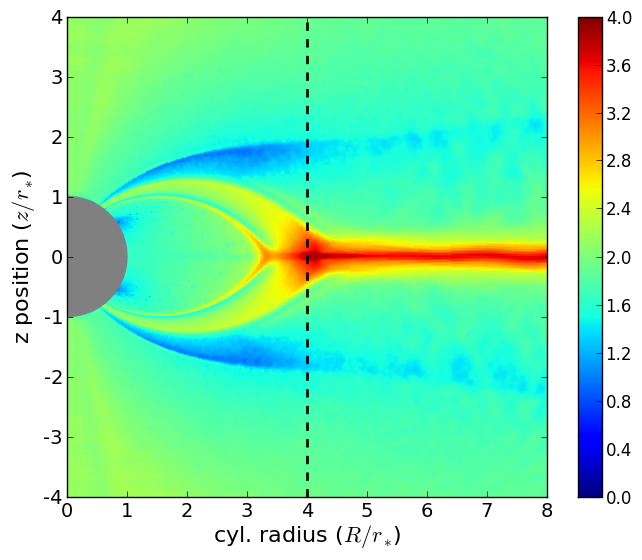}
		   \put(12.,87){\large \bf log particles per cell}
		  \put(79,79){\large \bf c)}
		  \end{overpic}}
\subfigure{\begin{overpic}
		  [width=.3233\textwidth]{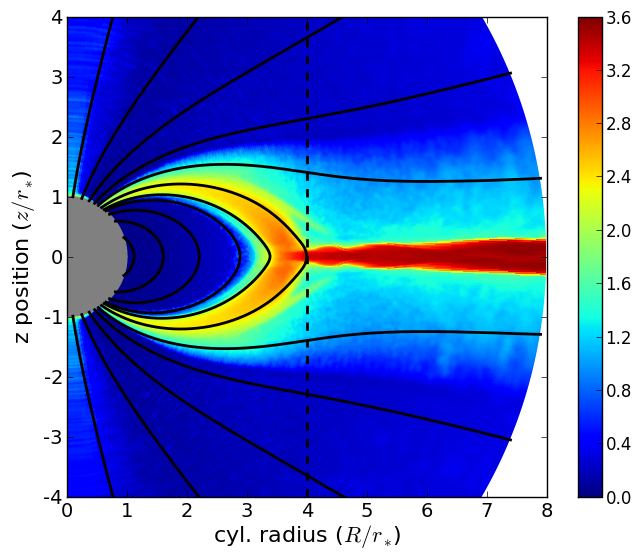}
		  \put(79,79){\large \bf d)}
		  \put(12,87){\large \bf log positron gamma}
		  \end{overpic}}
\subfigure{\begin{overpic}
		  [width=.3233\textwidth]{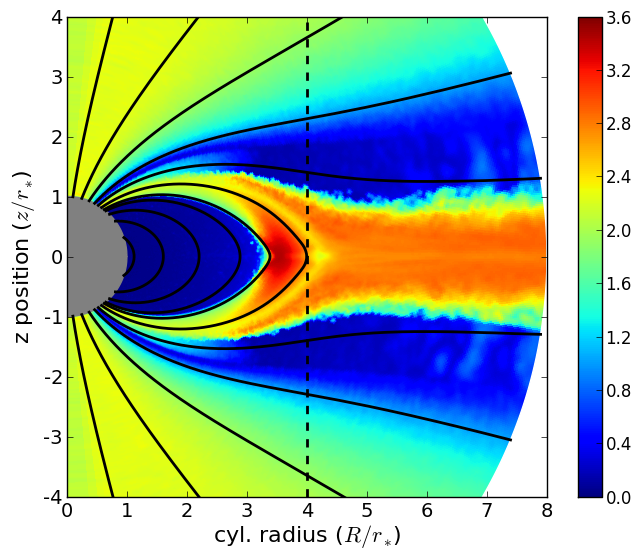}
		  \put(79,79){\large \bf e)}
		  \put(12,87){\large \bf log electron gamma}
		  \end{overpic}}
\subfigure{\begin{overpic}
		  [width=.3233\textwidth]{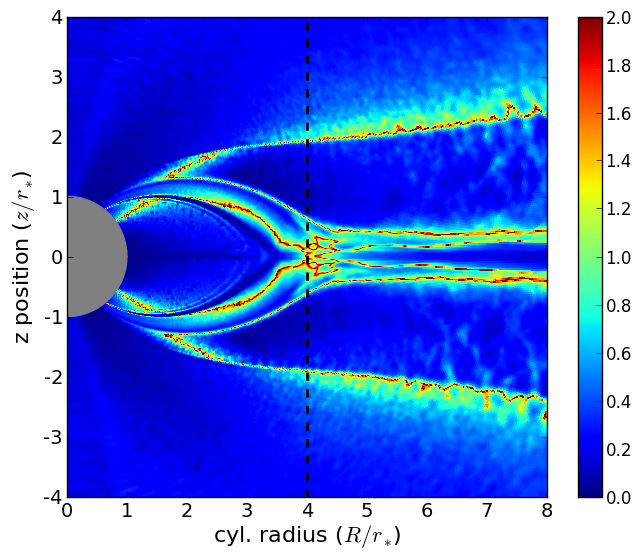}
		   \put(12,87){\large \bf log pair multiplicity}
		  \put(79,79){\large \bf f)}
		  \end{overpic}}
\caption{Simulation fvol\_.5\_rinj\_1.25 at $t=6P_*$. The neutron star is represented by the gray semicircle, the solid black curves are magnetic field lines, and the dashed vertical line is the light cylinder. a) $r^2J_m$ -- the sign is the same as that of $J_r$. b)  Log of the magnetization, $\sigma$. c) Log of the particles per cell (electrons and positrons combined). d/e) Log of the volume-averaged values of the positron/electron $\gamma$. f) Log of the pair multiplicity, $\kappa$. Sharp edges of high pair multiplicity are null surfaces.}
\label{canonicalfig}
\end{figure*}
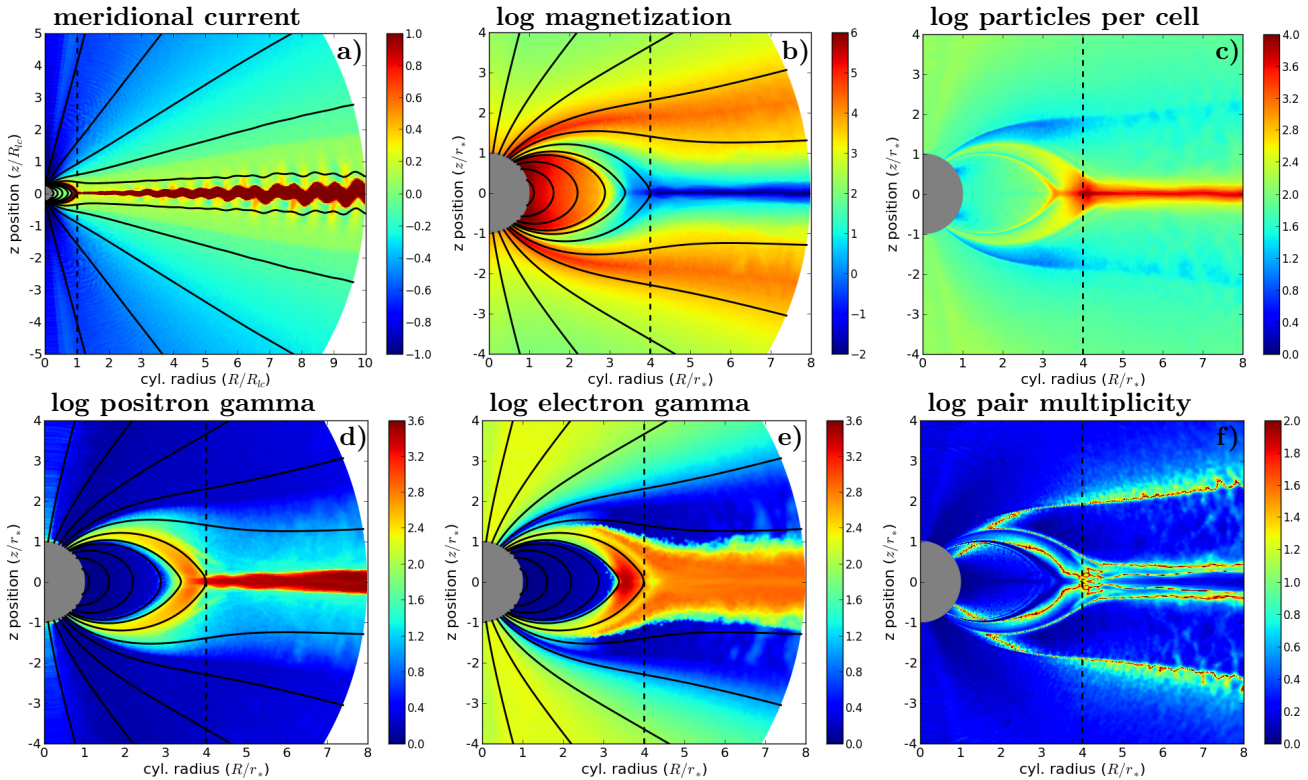

We begin by presenting general results that apply to all of our simulations which have volume injection turned on for the duration of the simulation. Fig. \ref{canonicalfig} shows snapshots of field values for a variety of physical quantities in simulation fvol\_.5\_rinj\_1.25 at $t=6P_*$.

Fig. \ref{canonicalfig}a shows $r^2J_m$, where $J_m$ is the current density in the meridional plane. The sign of $J_m$ is defined to be the same as the sign of the radial current density, $J_r$. The overall morphology of the current density is seen to be the same as that of the force-free solution. In particular, there is a negative radial current emanating from the polar cap region and a positive return current. The return current flows in a layer along the surface of the corotating torus of positively charged plasma bounded by the last open field line.

An important difference between our PIC solution and the force-free solution is that the equatorial current sheet in the PIC solution exhibits a kink instability beyond the light cylinder. Such a kink has also been observed in the simulations of \citet{PhilippovSpitkovsky} and \citet{CeruttiSpitkovsky}. The magnitude of the kink instability and even whether it is present or not depends on the parameter values of the simulation. In general, increasing the value of $f_\text{vol}$ while holding other parameters constant increases the strength of the kink instability. 

Fig. \ref{canonicalfig}b shows the logarithm of the magnetization, $\log(\sigma)$. Although the characteristic value of the magnetization at the light cylinder is high $\sigma \sim 10^2-10^4$, the magnetization drops to unity or below in the equatorial current layer and near the Y-point.

Fig. \ref{canonicalfig}c shows the logarithm of the total particles per cell (positrons and electrons combined). The characteristic particles per cell at the light cylinder is $N_\text{ppc} \approx 50$ outside the current layers (this is the particles per cell required for the Goldreich-Julian density) and between $N_\text{ppc} \approx 5 \times 10^3 - 5 \times 10^4$ inside the current layers. Note that due to logarithmic spacing in the radial direction, the cell volume at a given value of $\theta$ increases as $dV \propto r^3$, so the particles per cell actually grows as $\propto r$ beyond the light cylinder, since the density falls off as $r^{-2}$.

Figs. \ref{canonicalfig}d/e show the logarithm of the positron/electron gamma factors, respectively. The positrons have the highest gamma factors in strong current layers, and a sharp increase in the positron gamma factor is observed at the Y-point. This sharp increase is caused by dissipation at the Y-point, which results in a conversion of electromagnetic energy to particle energy (\S \ref{poyntfluxsec},\ref{dissipationsec}).

Fig. \ref{canonicalfig}f shows the logarithm of the pair multiplicity, defined as 
\begin{align}
\kappa \equiv \left| \frac{n_+ + n_-}{n_+ - n_-} \right|.
\end{align}
Due to computational constraints, our simulations are in the low pair multiplicity regime $\kappa \sim 1$ with the highest values of $\kappa$ achieved in the current layers inside the light cylinder. Sharp edges where $\kappa$ is large correspond to null surfaces where the charge density $\rho \equiv q(n_+ - n_-) = 0$.

\subsection{Spindown Luminosity}

\label{poyntfluxsec}
Having outlined the basic features of our PIC solutions, we now study the spindown luminosity of our solution and the transfer of electromagnetic energy to particles. 

The spindown luminosity for the aligned force-free rotator with a dipole magnetic field is given by the formula \citep{Spitkovsky}
\begin{align}
L_\text{ff} = c(r_* B_*)^2 \left(\frac{r_*}{R_\text{lc}}\right)^4.
\end{align}
The force-free solution has the Y-point at the light cylinder. However, since the spindown luminosity is determined by the amount of open magnetic flux, moving the Y-point inward/outward increases/decreases the amount of open magnetic flux and correspondingly increases/decreases the spindown luminosity relative to $L_\text{ff}$. 

\begin{figure}
\centering
\includegraphics[width=.49\textwidth]{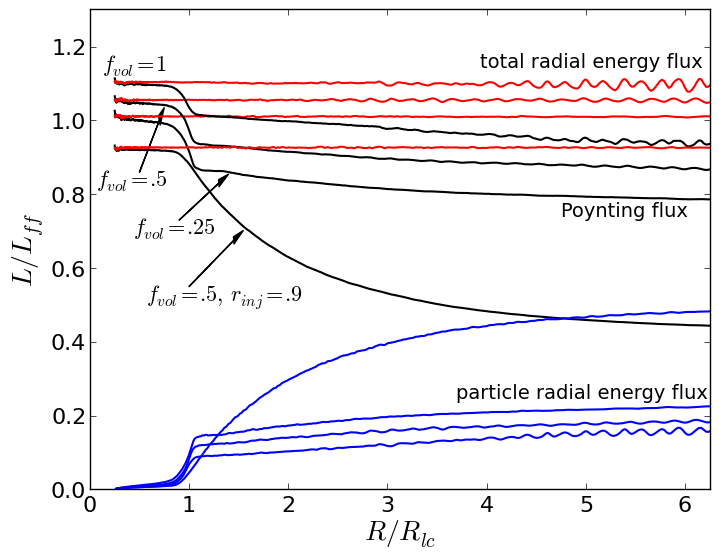}
\caption{Radial energy flux integrated over the surface of a sphere as a function of radius for simulations fvol\_.25\_rinj\_1.25, fvol\_.5\_rinj\_1.25, fvol\_1\_rinj\_1.25, and fvol\_.5\_rinj\_.9 at $t= 9P_*$. The black curves correspond to the radial component of the Poynting flux, the blue curves to the radial particle energy flux, and the red curves to the total radial energy flux (combination of the two).}
\label{triplepoyntfig}
\end{figure}

Fig. \ref{triplepoyntfig} shows the spindown luminosity as a function of radius at $t=9P_*$ for simulations (fvol\_.25\_rinj\_1.25, fvol\_.5\_rinj\_1.25, fvol\_.5\_rinj\_1.25\_Boris, fvol\_.5\_rinj\_.9). Different curves of the same color in each panel correspond to different values of the volume injection parameter $f_\text{vol}$. The curves of different colors show different components that make up the spindown luminosity. 

The black lines show the radial component of the Poynting flux integrated over the surface of a spherical shell at a given radius:
\begin{align}
L_\text{poynt}(r) \equiv \int r^2 d \Omega \bf S \cdot \rhat, 
\end{align}
where $\bfS \equiv c \bfE \times \bfB/ 4 \pi $ is the Poynting vector. The blue lines show the radial energy flux in particles at a given radius: 
\begin{align}
\label{prtlenergy}
L_\text{prtl}(r)  \equiv \frac{4 \pi r^2}{dV} \sum_{N(dV)} \gamma m c^2 v_r, 
\end{align}
where the sum is carried out over all particles (both electrons and positrons) inside a thin spherical shell of volume $dV$ centered on radius $r$. Because particles in the simulation are highly relativistic, it doesn't matter whether or not we subtract the particle rest mass energy (which is gained ``for free" when particles are injected) from the particle energy in equation (\ref{prtlenergy}). The red lines show the total spindown luminosity defined as $L_\text{tot} \equiv L_\text{poynt} + L_\text{prtl}$. 

We see from Fig. \ref{triplepoyntfig} that $L_\text{tot}$ is nearly constant in radius, and hence in time, meaning our solutions are in steady state. The ripples seen in some of the curves are associated with the kink instability (Fig. \ref{canonicalfig}a), which leads to a time-varying spindown luminosity. This effect is small, however, compared with the magnitude of the spindown luminosity. 

Simulations fvol\_.25\_rinj\_1.25, fvol\_.5\_rinj\_1.25, and fvol\_1\_rinj\_1.25 show that when there is sufficient plasma to screen the component of electric field along the magnetic field, $15-20\%$ of the force-free spindown luminosity carried by the Poynting flux is transferred to the particles, mostly in the vicinity of the light cylinder. However, in simulation fvol\_.5\_rinj\_.9, the maximum injection radius has been moved inside the light cylinder creating regions of $\bfE \cdot \bfB \ne 0$ in the outer magnetosphere. In this case, $\gtrsim 50\%$ of the spindown luminosity is transferred to the particles within $5 R_\text{lc}$, and the transfer of energy from the fields to the particles occurs over an extended region beyond the light cylinder. Note that we have also tried extending the injection out to $r_\text{inj} = 1.6 R_\text{lc}$ and see no significant difference compared to $r_\text{inj} = 1.25 R_\text{lc}$ for the same values of $f_\text{vol}$.

From simulations fvol\_.25\_rinj\_1.25, fvol\_.5\_rinj\_1.25, and fvol\_1\_rinj\_1.25, we see a trend that increasing the value of $f_\text{vol}$ tends to slightly increase the total spindown luminosity. This trend can be explained by the fact that simulations with a higher injection rate have more particles. Because the magnetization at the Y-point is low, particle inertia is important there and simulations with more particles will be able to force open a greater number of closed field lines driving the location of the Y-point inward and increasing the spindown luminosity. This effect is small, however, and the spindown luminosity is within $10 \%$ of the force-free value as long as particles are injected at the $E \times B$ drift velocity, and the Vay mover is used. 

Using the Boris mover, which does not accurately capture the $E \times B$ drift velocity when the Larmor radius is of order a cell \citep{BelyaevPIC}, or injecting particles with zero velocity rather than with the $E \times B$ drift velocity increases particle gamma factors in the simulation. This leads to a greater mass loading at the Y-point, driving it inward, and increasing the spindown luminosity compared to the force-free solution. Injecting particles with the $E \times B$ drift velocity and using the Vay mover, which we do in our simulations, minimizes the gyration of particles around magnetic field lines and allows them to $E \times B$ drift smoothly (at least initially). This situation is closest to reality, since a particle's Larmor radius should quickly be damped by synchrotron emission, which is not captured in our current simulations. 

\subsection{Dissipation and Transfer of EM Energy to Particles}
\label{dissipationsec}
\begin{figure}
\centering
\subfigure{\begin{overpic}[width=.4\textwidth]{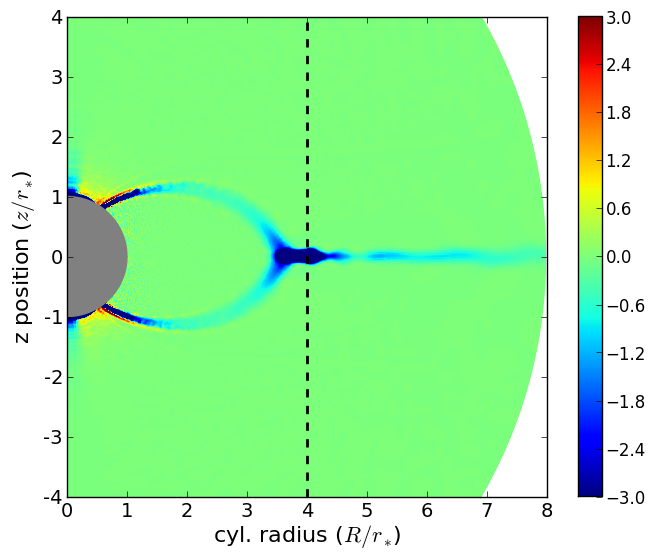}
		  \put(78,78){\large \bf a)}
		  \put(30,85){$f_\text{vol} = 1$,  $r_{inj} = 1.25 R_{lc} $}
		  \end{overpic}}
\subfigure{\begin{overpic}[width=.4\textwidth]{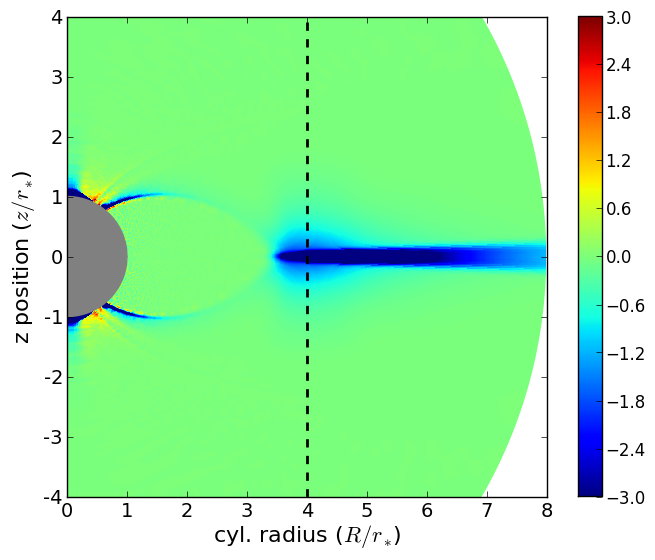}
		  \put(78,78){\large \bf b)}
		  \put(30,85){$f_\text{vol} = .5$,  $r_{inj} = .9 R_{lc}$}
		  \end{overpic}}
\caption{$- \bfE \cdot \bfJ/(u_\text{lc} \Omega_*)$ for simulations fvol\_1\_rinj\_1.25 (panel a) and fvol\_.5\_rinj\_.9 (panel b) at $t=6P_*$. Negative regions correspond to dissipation of Poynting flux and conversion of electromagnetic energy to particle energy.}
\label{EdotJfig}
\end{figure}

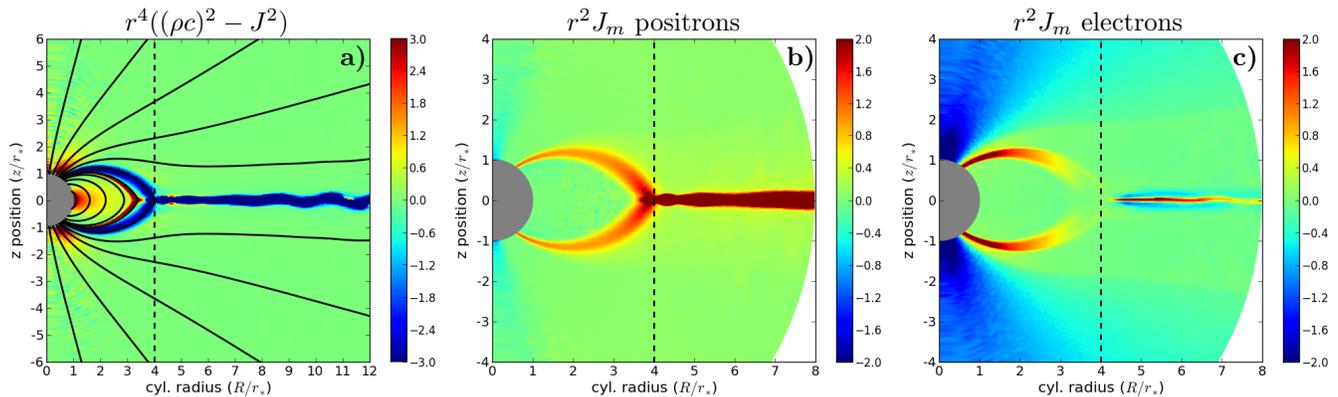
\begin{figure*}
\centering
\subfigure{\begin{overpic}[width=.33\textwidth]{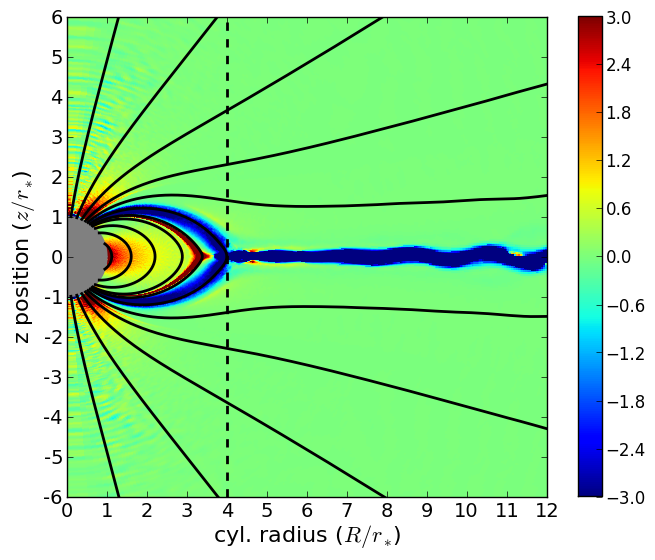}
		  \put(77,77){\large \bf a)}
		  \put(28,85){\large $r^4((\rho c)^2 - J^2)$}
		  \end{overpic}}
\subfigure{\begin{overpic}[width=.33\textwidth]{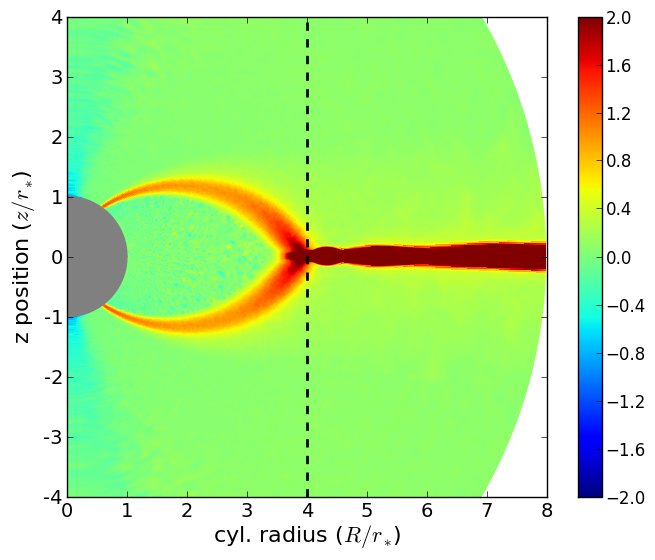}
		  \put(77,77){\large \bf b)}
		  \put(27,85){\large $r^2 J_m$ positrons}
		  \end{overpic}}
\subfigure{\begin{overpic}[width=.33\textwidth]{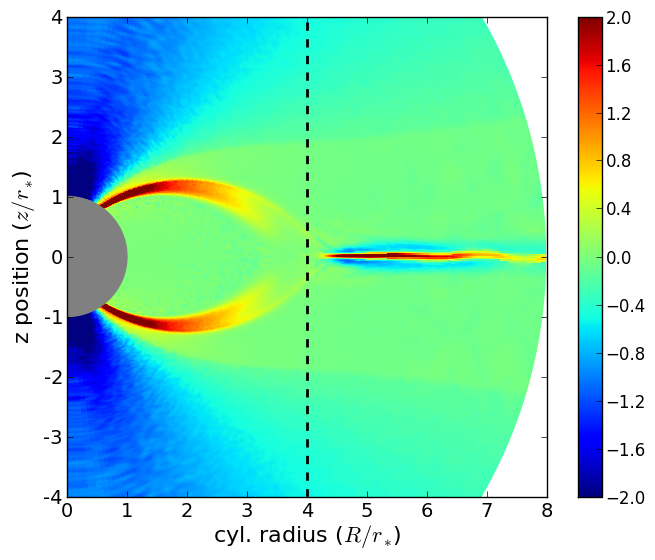}
		  \put(77,77){\large \bf c)}
		  \put(27,85){\large $r^2 J_m$ electrons}
		  \end{overpic}}
\caption{Simulation fvol\_1\_rinj\_1.25 at $t = 6 P_*$. a) Radially-weighted square of the magnitude of the current four vector. b) Radially-weighted current in the meridional plane for the positrons c) Radially weighted current in the meridional plane for the electrons.}
\label{spacelikefig}
\end{figure*}

To obtain a clearer picture of where the electromagnetic energy is converted to particle energy in the magnetosphere, we can look at $\bfE \cdot \bfJ$, which determines the rate of dissipation of electromagnetic energy according to Poynting's theorem:
\begin{align}
\frac{\partial u}{\partial t} + \bfnabla \cdot \bfS = - \bfE \cdot \bfJ.
\end{align}

Panels a and b of Fig. \ref{EdotJfig} show the value of $-\bfE \cdot \bfJ/(u_\text{lc} \Omega_*)$ for simulations fvol\_1\_rinj\_1.25 and fvol\_.5\_rinj\_.9, respectively at $t=6P_*$. Here, $u_\text{lc} \equiv (B_*^2/4 \pi) (r_*/R_\text{lc})^6$ is the characteristic value of the electromagnetic energy density at the light cylinder. The regions that have negative $-\bfE \cdot \bfJ$ are dissipative regions, where the electromagnetic Poynting flux is converted to particle energy. 

We see in Fig. \ref{EdotJfig} that there are three distinct regions, where dissipation is important. These are a low altitude region above the polar cap, the current layers that bound the torus, and a region around the Y-point, extending into the equatorial current sheet.

Comparing panels a and b of Fig. \ref{EdotJfig}, we see that the dissipative region around the Y-point extends into the current sheet in panel b, but is more pronounced in the direct vicinity of the Y-point in panel a. This is due to a better screening of the accelerating electric field in the current sheet in simulation fvol\_1\_rinj\_1.25 compared to simulation fvol\_.5\_rinj\_.9, for which the cutoff radius for volume injection is inside the light cylinder at $r_\text{inj} = .9 R_\text{lc}$. This also explains why the conversion of Poynting flux to radial particle energy flux in Fig. \ref{triplepoyntfig} extends over a larger radius for simulation fvol\_.5\_rinj\_.9 compared to those simulations which have $r_\text{inj} = 1.25 R_\text{lc}$.

\subsection{Current Flow Within the Magnetosphere}

Another key result of our simulations is the spacelike nature of the current layers. In regions of spacelike current, $(\rho c)^2 - J^2 < 0$, and there is necessarily counterstreaming of oppositely charged particles. Fig. \ref{spacelikefig}a shows $r^4((\rho c)^2 - J^2)$, normalized by a canonical value, for simulation fvol\_.5\_rinj\_1.25 at $t = 6P_*$. Negative regions of the figure correspond to a spacelike current density, positive regions are ones where particles are non-relativistic (at least on average), and null regions are ones where the current is carried by a single species traveling relativistically.

The fact that the current layers are spacelike implies counterstreaming, but does not automatically imply that the electrons and positrons flow in opposite directions relative to the neutron star. To demonstrate the latter explicitly, we show the current in the meridional plane weighted by the square of the radius, $r^2 J_m$, due to positrons and electrons separately in panels b and c of Fig. \ref{spacelikefig}.  As in Fig. \ref{canonicalfig}a, the sign of $J_m$ is defined to be the same as the sign of the radial current density, $J_r$. 

From the two panels, we immediately see that both electrons and positrons provide a positive contribution to the current in the current layers. Thus, the electrons and positrons are moving in opposite directions with respect to the neutron star. In the case of the aligned rotator, which we are simulating, the positrons flow away from the star and the electrons fall back on the star. In the case where the spin and magnetic axes are anti-aligned, the current densities would be the same, but the roles of the electrons and the positrons would be reversed, so electrons would flow outwards and positrons inwards.

\subsection{Turning Off the Force-Free Solution}

\begin{figure*}
\subfigure{\begin{overpic}[width=.32\textwidth]{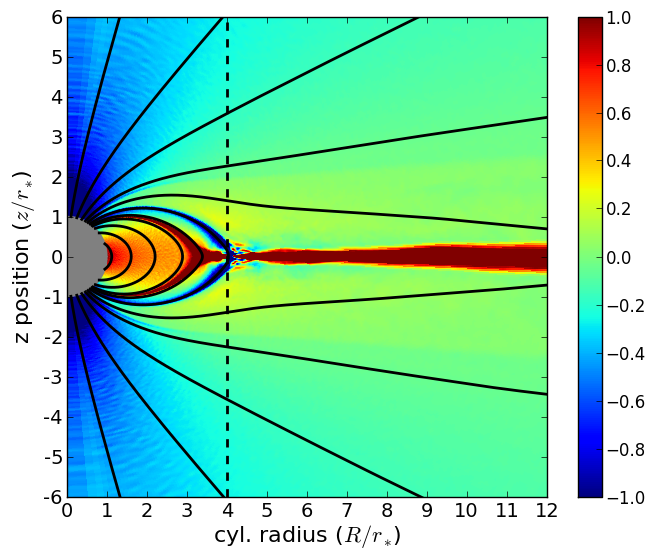}
		  \put(77,77){\large \bf a)}
		  \put(27,85){$r^2 \rho$ at $t=3P_*$}
		  \end{overpic}}
\subfigure{\begin{overpic}[width=.32\textwidth]{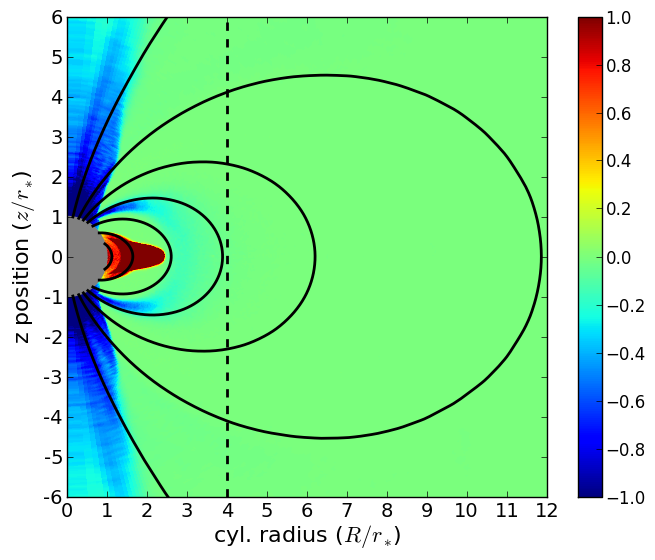}
		  \put(77,77){\large \bf b)}
		  \put(27,85){$r^2 \rho$ at $t=8P_*$}
		  \end{overpic}}
\subfigure{\begin{overpic}[width=.35\textwidth]{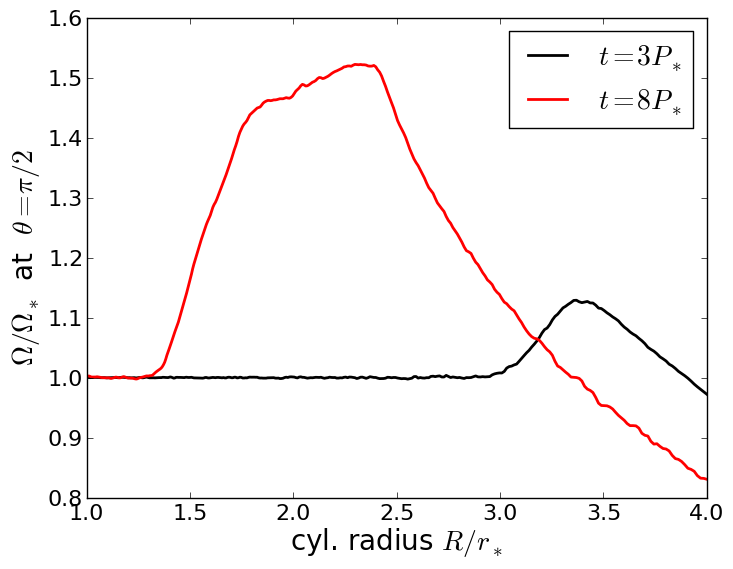}
		  \put(15,70){\large \bf c)}
		  \put(27,77){rotation profile at equator}
		  \end{overpic}}
\centering
\caption{Simulation  injvol\_turnoff. a) Radially-weighted charge density at $t=3P_*$, before volume injection has been turned off. b) Radially-weighted charge density at $t=8P_*$, 5 rotational periods after volume injection has been turned off. c) Rotation profile in the equatorial plane for the force-free solution ($t=3P_*$) and the dome-disk solution $(t=8P_*$)}
\label{turnoff_fig}
\end{figure*}
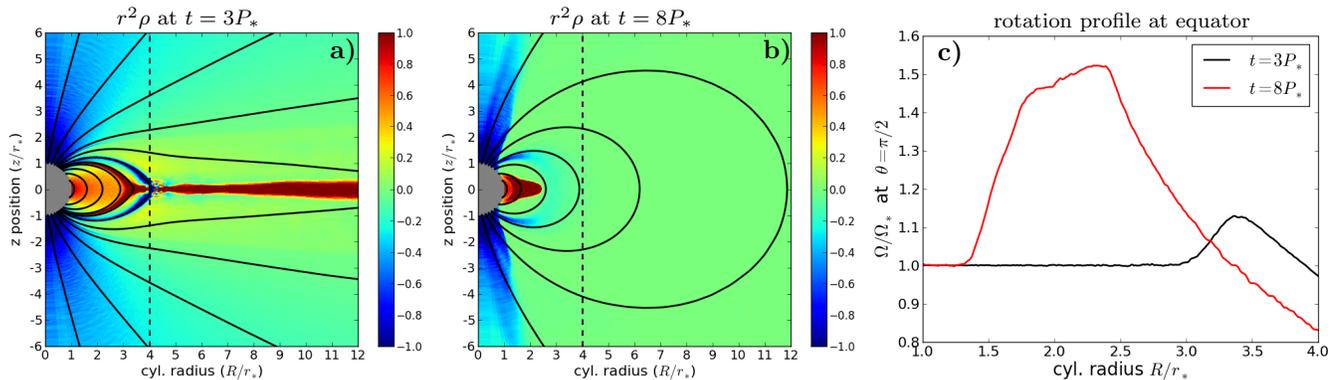

An interesting consequence of a spacelike current layer is that it is not possible to maintain a force-free magnetosphere with only emission of surface charge from the star. This is because a spacelike current density implies that the current is greater than what can be supplied by the Goldreich-Julian density traveling at the speed of light. To show this point explicitly, we turn on volume injection for $3 P_*$, and then turn it off for the remainder of the simulation, leaving only surface emission (simulation injvol\_turnoff).

Panels a and b of Fig. \ref{turnoff_fig} show the charge density weighted by the square of the radius, $r^2 \rho$, normalized by a canonical value at $t=3P_*$, before the volume injection is turned off, and at $t=8P_*$, after it has been turned off for $5P_*$. In panel a, we see that the magnetosphere resembles the force-free case and is filled with plasma. 

However, we see in panel b that after volume injection has been turned off for five rotational periods of the star, large vacuum gaps have opened up in the outer magnetosphere. There is no longer any appreciable current flowing the system, so the magnetic field structure has collapsed to the original dipolar configuration. This is the dead ``electrosphere" of \citep{MichelLi} also known as the ``dome-disk" solution for the domes of trapped electrons and disky torus of trapped positrons. Thus, even if a force-free state is realized by one means or another, it cannot be sustained with injection of the surface charge from the pulsar alone. 

As an independent way of verifying the code, it is interesting to consider the rotation profile in the equatorial plane for simulation injvol\_turnoff at $t=3P_*$ and $t=8P_*$ (panel c of Fig. \ref{turnoff_fig}). This rotation profile is generated from the fields using the $\phi$-component of $c\bfE \times \bfB/B^2$ at the equator. 

From the rotation profile at $t=3P_*$, which corresponds to the force-free state, we see that corotation electric and magnetic fields extend out to $\approx 3r_* = .75 R_\text{lc}$, at which point we approach the $Y$-point region and there is a $10 \%$ deviation from corotation. At $t=8P_*$, after volume injection has been turned off and the force-free solution has collapsed to the dome-disk solution, we see that corotation electric and magnetic fields only extend to approximately $1.3 r_*$ at the equator. 

This is explained by Ferraro's isorotation theorem which states that only particles on field lines completely immersed in plasma are required to corotate with the star. Because the dome-disk solution contains large vacuum gaps in the outer magnetosphere, field lines which cross the equator beyond $1.3 r_*$ pass through a vacuum region, so particles on these field lines will generally not corotate with the star.
 
\section{Discussion}
\label{discussion}
We have performed axisymmetric PIC simulations of the pulsar magnetosphere. An important advantage of our simulations over resistive simulations (e.g. resisitive force-free or MHD) is that dissipation is handled self-consistently. This is by virtue of the PIC method, which is capable of directly simulating microphysical plasma processes, as well as transfer of electromagnetic energy to the particles.

We find that dissipation and particle acceleration occur in the current sheets and at the Y-point. The strongest dissipation occurs near the Y-point, and at least $15-20 \%$ of the electromagnetic spindown luminosity is transferred to the particles within 5 light cylinder radii. The amount of dissipation can be substantially higher, $\gtrsim 50\%$, if there is not enough plasma in the outer magnetosphere to screen the accelerating electric field. In this case, the region of strong dissipation is not confined to the Y-point and extends to several light cylinder radii. These high-dissipation solutions can be astrophysically relevant as up to an order unity fraction of the spindown luminosity is released in gamma rays for the millisecond pulsars observed by Fermi \citep{Fermi}.

In addition to the dissipation occuring at the Y-point and in the current sheets due to reconnection and the presence of an accelerating electric field, plasma instabilities are another possible source of reconnection. The two relevant instabilities for pulsars are relativistic tearing and kinking modes \citep{ZenitaniHoshino2007,ZenitaniHoshino2008}. Because our simulations are axisymmetric, we are unable to observe the tearing mode, but we do observe the kinking mode. The kink instability in our simulations leads to corrugation of the current sheet, without disrupting it. If the current sheet were disrupted due to the kink instability, this would lead to greater particle acceleration and dissipation in the equatorial current sheet beyond the light cylinder. Additionally, if plasmoids were periodically emitted from the Y-point, which is not seen in our simulations, but was observed by \citet{chenbeloborodov}, then one would expect a significant temporal variation for the dissipation in the magnetosphere. 

Other than dissipation in the current sheets and at the Y-point, our solution is morphologically similar to the force-free solution. In particular, the value of the spindown luminosity is within $10 \%$ of the force-free value when the component of electric field parallel to the magnetic field is well-screened out to the light cylinder. However, we find that due to spacelike current sheets in the force-free solution, a force-free state cannot be supported simply by pulling off the surface charge from the star.

This does not necessarily mean that models of pulsars having only low-altitude pair production are not viable. \citet{CeruttiSpitkovsky} have shown that if a high magnetization is maintained at the surface of the neutron star and particles are given an outward kick, encouraging them to flow out into the magnetosphere, then surface production of particles is sufficient to fill the magnetospheric vacuum gaps with plasma and generate a force-free solution. Ultimately, simulations implementing more detailed models of the pair production physics will be required to determine the conditions under which gaps are present or absent in the magnetosphere.

\section*{Acknowledgements}
The author would like to thank Anatoly Spitkovsky for inspiring this work, Jon Arons and Sasha Philippov for stimulating discussions, and the referee, Alexander Chen, for his insightful comments, which greatly helped improve the paper. The research described here was partially supported by NASA Astrophysics Theory grant NNX14AH49G to the University of California, Berkeley. This work used the Extreme Science and Engineering Discovery Environment (XSEDE), which is supported by National Science Foundation grant number ACI-1053575 \citep{XSEDE}. This research also used the SAVIO computational cluster resource provided by the Berkeley Research Computing program at the University of California Berkeley (Supported by UC Chancellor, UC Berkeley Vice Chancellor of Research and Office of the CIO). 

\bibliography{../bibliography/pulsar}

\appendix

\end{document}